\documentclass[reprint,superscriptaddress,
preprintnumbers,
 amsmath,amssymb,
 aps,
]{revtex4-2}

\usepackage{graphicx}
\usepackage{dcolumn}
\usepackage{bm}
\usepackage{xcolor}
\usepackage{subcaption}
\usepackage{ulem}
\usepackage{multirow}
\usepackage{url}

\def\N{{$^{13}${\rm N}} }

\def\reac{{$^{13}${N($\alpha$,$p$)}$^{16}${O}} }

\def\ap{($\alpha$,$p$) }
\def\aa{($\alpha$,$\alpha$) }
\def\Nap{$^{13}$N($\alpha$,$p$)$^{16}$O }
\def\Opa{$^{16}$O($p$,$\alpha$)$^{13}$N }
\def\Naa{$^{13}$N($\alpha$,$\alpha$)$^{13}$N }
\def\ea{\textit{et al.} }

\begin{document}

\preprint{APS/123-QED}

\title{First direct measurement of the \boldmath{\reac} reaction relevant for core-collapse supernovae nucleosynthesis}

\author{H. Jayatissa}
\email{hjayatissa@anl.gov}
\affiliation{Physics Division, Argonne National Laboratory, Lemont, IL, 60439, USA}
\author{M.~L.~Avila}
\affiliation{Physics Division, Argonne National Laboratory, Lemont, IL, 60439, USA}
\author{K. E. Rehm}
\affiliation{Physics Division, Argonne National Laboratory, Lemont, IL, 60439, USA}
\author{R. Talwar}
\affiliation{Physics Division, Argonne National Laboratory, Lemont, IL, 60439, USA}
\author{P. Mohr}
\affiliation{Institute for Nuclear Research (Atomki), P.O. Box 51, Debrecen H-4001, Hungary}
\author{K. Auranen}
\altaffiliation{University of Jyvaskyla, Department of Physics, P.O. Box 35, FI-40014 University of Jyvaskyla, Finland}
\affiliation{Physics Division, Argonne National Laboratory, Lemont, IL, 60439, USA}
\author{J. Chen}
\affiliation{Physics Division, Argonne National Laboratory, Lemont, IL, 60439, USA}
\author{D.~A.~Gorelov}
\affiliation{Department of Physics and Astronomy, University of Manitoba, Winnipeg, Manitoba R3T 2N2, Canada}
\affiliation{Physics Division, Argonne National Laboratory, Lemont, IL, 60439, USA}
\author{C. R. Hoffman}
\affiliation{Physics Division, Argonne National Laboratory, Lemont, IL, 60439, USA}
\author{C. L. Jiang}
\affiliation{Physics Division, Argonne National Laboratory, Lemont, IL, 60439, USA}
\author{B. P. Kay}
\affiliation{Physics Division, Argonne National Laboratory, Lemont, IL, 60439, USA}
\author{S. A. Kuvin}
\altaffiliation{Los Alamos National Laboratory, Los Alamos, New Mexico 87545, USA}
\affiliation{Physics Division, Argonne National Laboratory, Lemont, IL, 60439, USA}
\author{D. Santiago-Gonzalez}
\affiliation{Physics Division, Argonne National Laboratory, Lemont, IL, 60439, USA}




\date{\today}

\begin{abstract}

Understanding the explosion mechanism of a core-collapse supernova (CCSN) is important to accurately model CCSN scenarios for different progenitor stars using model-observation comparisons. The uncertainties of various nuclear reaction rates relevant for CCSN scenarios strongly affect the accuracy of these stellar models. Out of these reactions, the \Nap reaction has been found to affect various stages of a CCSN at varying temperatures. This work presents the first direct measurement of the \Nap reaction performed using a 34.6 MeV beam of radioactive \N ions and the active-target detector MUSIC (MUlti-Sampling Ionization Chamber) at Argonne National Laboratory. The resulting total \Nap reaction cross sections from this measurement in the center-of-mass energy range of 3.26 - 6.02 MeV are presented and compared with calculations using the Hauser-Feshbach formalism. Uncertainties in the reaction rate have been dramatically reduced at CCSN temperatures.

\end{abstract}

\maketitle




Supernova explosions are important sites for the nucleosynthesis of chemical elements \cite{Woosley1995,Rauscher2002,Hix2014,Chieffi2017,Curtis2019,Arnett2020}. Core-collapse supernovae (CCSNe) occur when massive stars ($M>$ 8 $M_{\odot}$) exhaust their fuel in the core, resulting in the gravitational collapse of the iron core \cite{Woosley2005}. When the density of the core reaches nuclear matter density, the repulsive nuclear forces create an outward shock wave that results in one of the strongest explosions in the universe, ejecting a variety of chemical elements into the interstellar medium. Properties of CCSNe can be obtained by studying the signatures from prominent remnants such as $^{44}$Ti and $^{56}$Ni \cite{Young2006,Young2007,Fryer2012,Deihl2021}. A large number of nuclear reactions affect the production of these isotopes and precise knowledge of nuclear reaction rates are needed to constrain astrophysical models and to obtain accurate information about the CCSN \cite{Fryer2018}. 
Several sensitivity studies have been performed throughout the years to identify critical reactions that affect the final composition of CCSN nucleosynthesis \cite{The1998,Hoffman1999,Jordan2003,Magkotsios2010}. A recent sensitivity study was performed by Subedi \ea \cite{Subedi2020}, in which the rates of various reactions were varied and their impact on the synthesis of $^{44}$Ti and $^{56}$Ni isotopes were inferred. In these calculations, a 1-D model was evolved for 15 $\sim M_{\odot}$, 18 $\sim M_{\odot}$ and 22 $\sim M_{\odot}$ progenitor stars from zero-age main sequence through the explosion. Their work has identified the \Nap reaction as one of 18 reactions that significantly impact the abundances of $^{44}$Ti and $^{56}$Ni, as well as the ratio between the two isotopes. Rate variation factors of 10 and 100, depending on the existing experimental and theoretical data were explored. For the case of the \Nap reaction, the currently available reaction rate from the Caughlan and Fowler (CF88) compilation \cite{CF88} published in REACLIB \cite{Reaclib} is based on the time-inverse reaction \Opa, and the use of the Hauser-Feshbach formalism is not considered to be valid due to the low level density in the compound nucleus $^{17}$F. Due to the large uncertainty and lack of information available on this reaction, a factor of 100 rate variation was used, revealing that the CCSN yield of the $^{44}$Ti and $^{56}$Ni isotopes decreases significantly by increasing the \Nap reaction rate within the temperature range of 1.9 -- 6.2 GK.

Another recent sensitivity study by Hermansen \ea \cite{Hermansen2020} used a 1-D explosive silicon burning model for CCSN environments. Using the current uncertainties in the STARLIB reaction rate library \cite{starlib}, they identified 48 reactions for a 12~M$_{\odot}$ progenitor star that significantly influence the production of long-lived radioisotopes. Again, the \Nap reaction has been identified as specifically affecting the production of $^{44}$Ti, $^{48,49}$V, $^{51}$Cr, $^{52,53}$Mn and $^{55}$Fe isotopes produced in CCSN. This reaction was found to be one of the bottlenecks in the buildup of heavy elements during nuclear statistical equilibrium freeze-out. The temperature range where the reaction rate needs to be constrained to reliably predict nucleosynthesis ranges from $\approx$ 0.75 to $\approx$ 5.6~GK.


The \Nap reaction can also affect the amount of $^{13}$C observed in presolar SiC grains from CCSN by reducing the amount of available \N produced via the hot CNO cycle. Pignatari \ea \cite{Pignatari2015} have suggested that ingestion of hydrogen into the helium shell of massive stars during the shock propagation of CCSN explosions allows proton capture on the available $^{12}$C to create an excess of \N. This could possibly explain the high yields of $^{13}$C observed in presolar grains compared to the solar composition. During the supernova, the $^{13}$C production is thus affected in a temperature region of $\leq$1 GK during the supernova shock propagation. The reaction rate of the inverse reaction \Opa also plays a role in the creation of $^{12}$C by oxygen burning at high proton abundances via $^{16}$O(p,$\alpha$)$^{13}$N($\gamma$,p)$^{12}$C. This in turn affects the abundances of argon and calcium in type Ia supernovae nucleosynthesis.
 

The \Nap reaction cross section has not been measured directly in the past. Various other reaction mechanisms have been used in order to infer its reaction rate for astrophysical interest. The presently available \Nap reaction rate from the CF88 compilation \cite{CF88} is obtained using the cross-section measurements of the inverse $^{16}$O(p,$\alpha$)\N reaction and the detailed balance theorem. No associated uncertainties for this reaction rate are given and very little information on the data is available. In addition, this rate only constrains the contribution from the ground state of $^{16}$O which might not be accurate for reaction rates at the high temperatures relevant for CCSN where excited states in $^{16}$O are expected to have a significant contribution to the reaction rate. 

A recent attempt at obtaining the \Nap reaction rate was performed by A. Meyer \ea \cite{Meyer2020} by studying the unbound states of the compound nucleus $^{17}$F by measuring relevant states of the isobaric analog $^{17}$O, using the $^{13}$C($^7$Li,$t$)$^{17}$O reaction. Their analysis is hindered by the lack of information regarding the partial $\alpha$ widths for the relevant $^{17}$F states that have known analog states in $^{17}$O. As such, the focus is mainly on low-lying resonances in $^{17}$F relevant for the \Nap reaction at temperatures below 1.4 GK. For higher temperatures, Ref.~\cite{Meyer2020} normalized the Hauser-Feshbach rate given in STARLIB \cite{starlib}. This reaction rate at temperatures relevant for CCSN is up to a factor of 6 higher than the REACLIB rate \cite{Reaclib}. This discrepancy emphasizes significant uncertainties for the \Nap reaction rate, highlighting the importance of a direct measurement.

The present paper reports the first direct measurement of the \Nap reaction cross sections in the center-of-mass energy range of 3.26 - 6.02 MeV in order to infer the \Nap reaction rate relevant for CCSN.


The first direct measurement of the \Nap reaction was carried out at the Argonne Tandem Linac Accelerator System (ATLAS) at Argonne National Laboratory. A radioactive $^{13}$N$^{7+}$ beam was created with a 50 MeV $^{12}$C$^{5+}$ primary beam using the $^{12}$C($d,n$)\N reaction via the in-flight technique \cite{Harss00}. The maximum \N beam intensity was around 1000 pps with an approximate purity of 50 \%. The energy of the $^{13}$N$^{7+}$ beam was determined using the magnetic rigidity of the beam passing through a bending magnet located upstream of the target. This magnet is used for separating the \N beam from the primary $^{12}$C beam with the field settings calibrated from previous stable beam measurements for which the energies have been measured using the ATLAS time-of-flight system \cite{PARDO1988226}. The beam energy of the secondary \N was calculated to be 34.6 $\pm$ 0.7 MeV.  

The \Nap reaction cross section was measured using the MUlti-Sampling Ionization Chamber (MUSIC) detector \cite{MUSIC_NIM}. The anode is segmented in 18 strips, each with a width of 15.78 mm. The 16 center strips are subdivided in asymmetric left and right sections. More details of the MUSIC detector and segmentation of the anode can be found in Ref.~\cite{MUSIC_NIM}. Due to the structure of the segmented anode pad of the MUSIC detector and because the \N beam loses energy as it travels through the gas volume of the detector, each anode strip can be used as a separate energy data point in an excitation function covering a large energy range using one incident beam energy. The energy binning size of each point is determined by the amount of energy lost by the beam in the width of each anode strip. The MUSIC detector chamber consist of beam entrance and exit windows made of 1.3 mg/cm$^2$ Ti. There is a 35.9 mm dead layer between the entrance window and the first anode strip (strip 0). The MUSIC detector was filled with a He-Kr (95\% - 5\% by volume) gas mixture. The pressure inside the MUSIC chamber was measured to be 402 Torr.

\begin{figure}[h]
\includegraphics[width=0.5\textwidth]{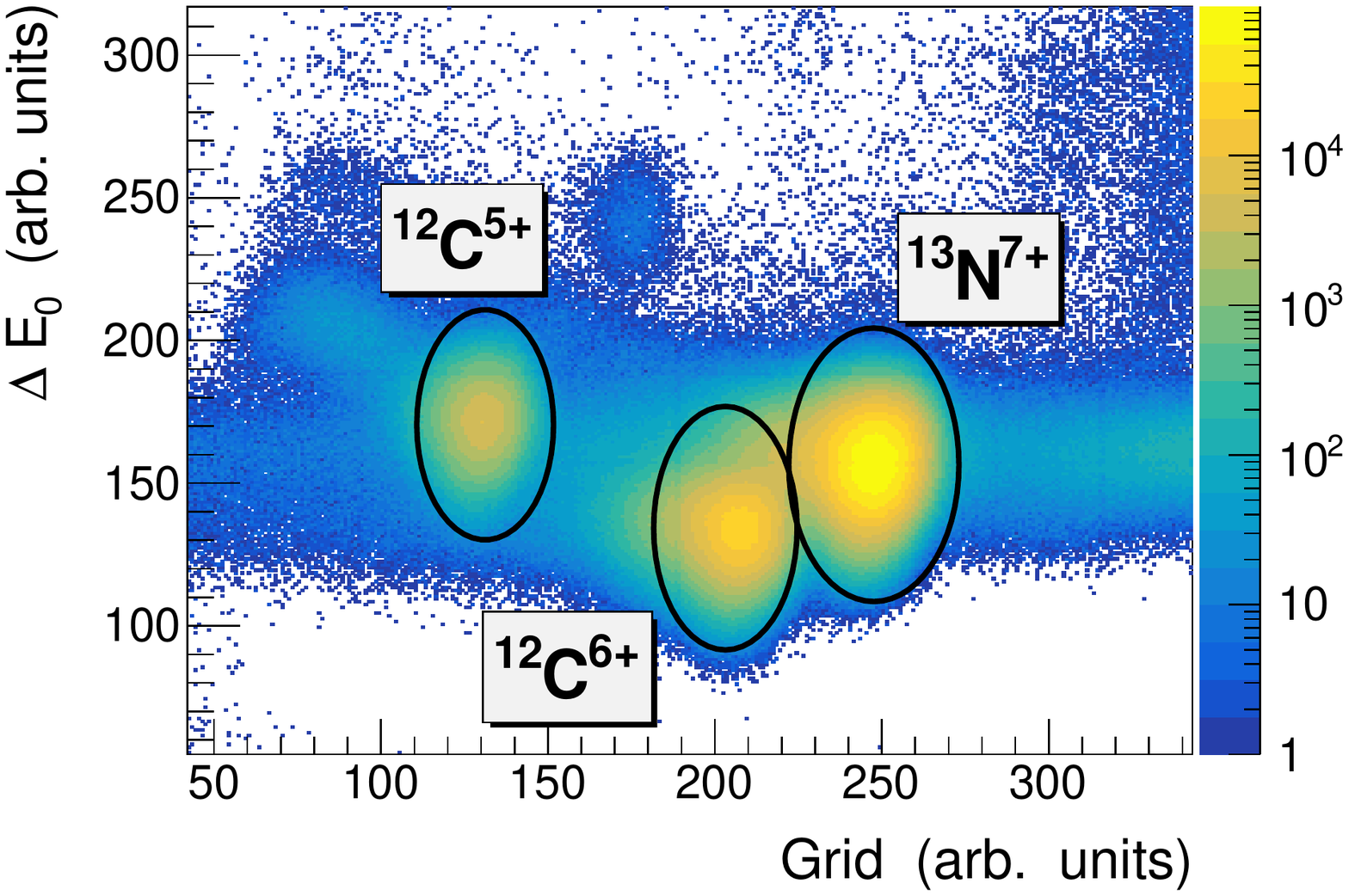}
\caption{\label{fig:grid_stp0} Plot of the Grid vs Strip0 showing the $^{13}$N$^{7+}$ beam and contaminants from the primary $^{12}$C beam.}
\end{figure}


The first anode strip (strip 0) and the signal from the Frisch grid are used to identify the beam for normalization purposes. The \N beam was identified from the main contaminants (different charge states of the primary  $^{12}$C beam) using the energy deposited in the grid and Strip 0, as shown in Fig. \ref{fig:grid_stp0}. An advantage of MUSIC is that it allows for self normalization of the absolute cross section by counting the total number of \N beam particles that entered the gas volume.

\begin{figure}[ht]
\includegraphics[width=0.5\textwidth]{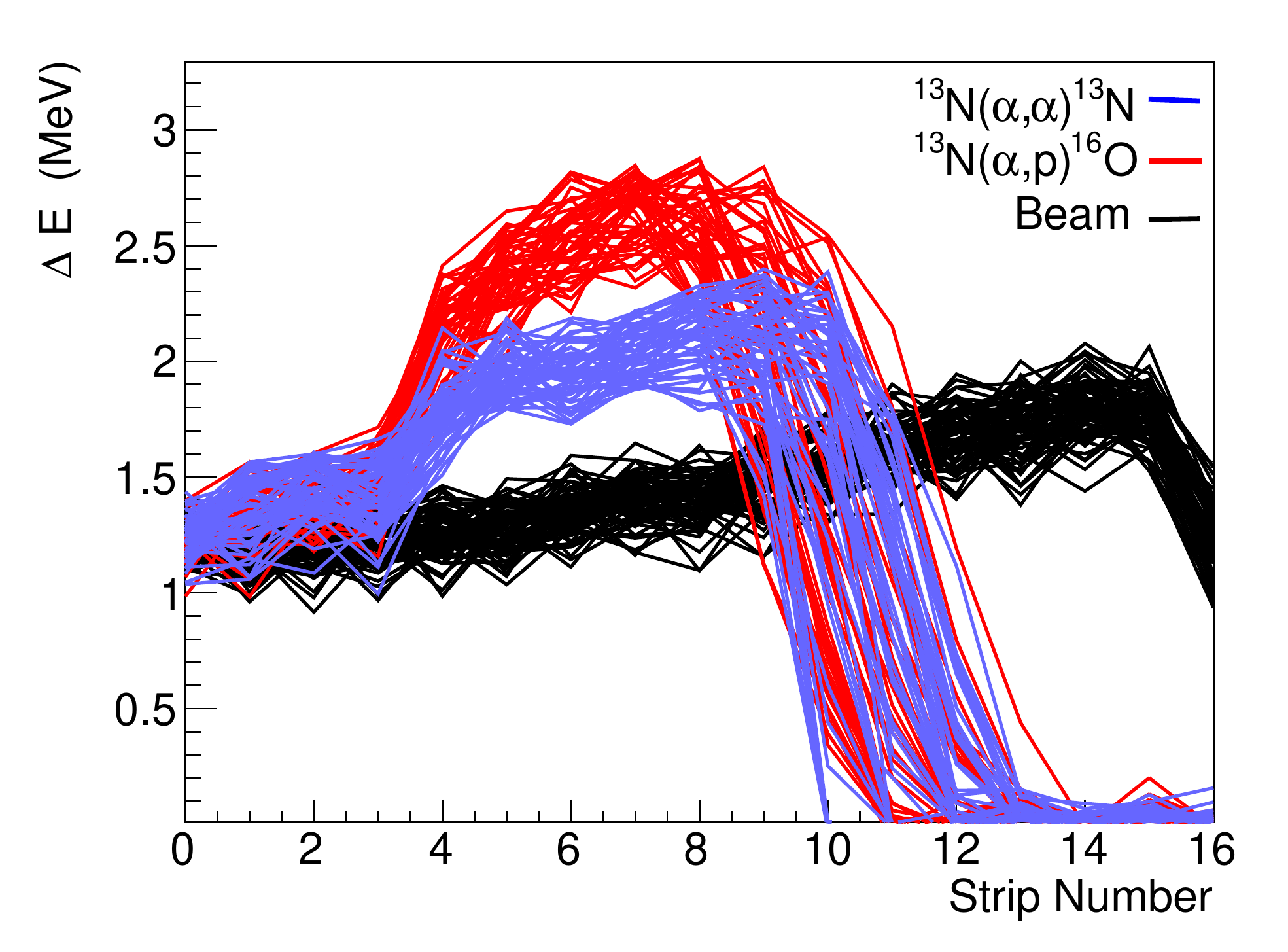}
\caption{\label{fig:traces}Energy loss per anode strip of different reaction channels in MUSIC from the present work. Shown are the energy losses for the unreacted  \N beam (black), \Naa events occurring in anode strip 4 (blue) and \Nap events occurring in anode strip 4 (red).}
\end{figure}

\begin{figure*}
\begin{subfigure}{.33\textwidth}
  \centering
  \includegraphics[width=1.0\linewidth]{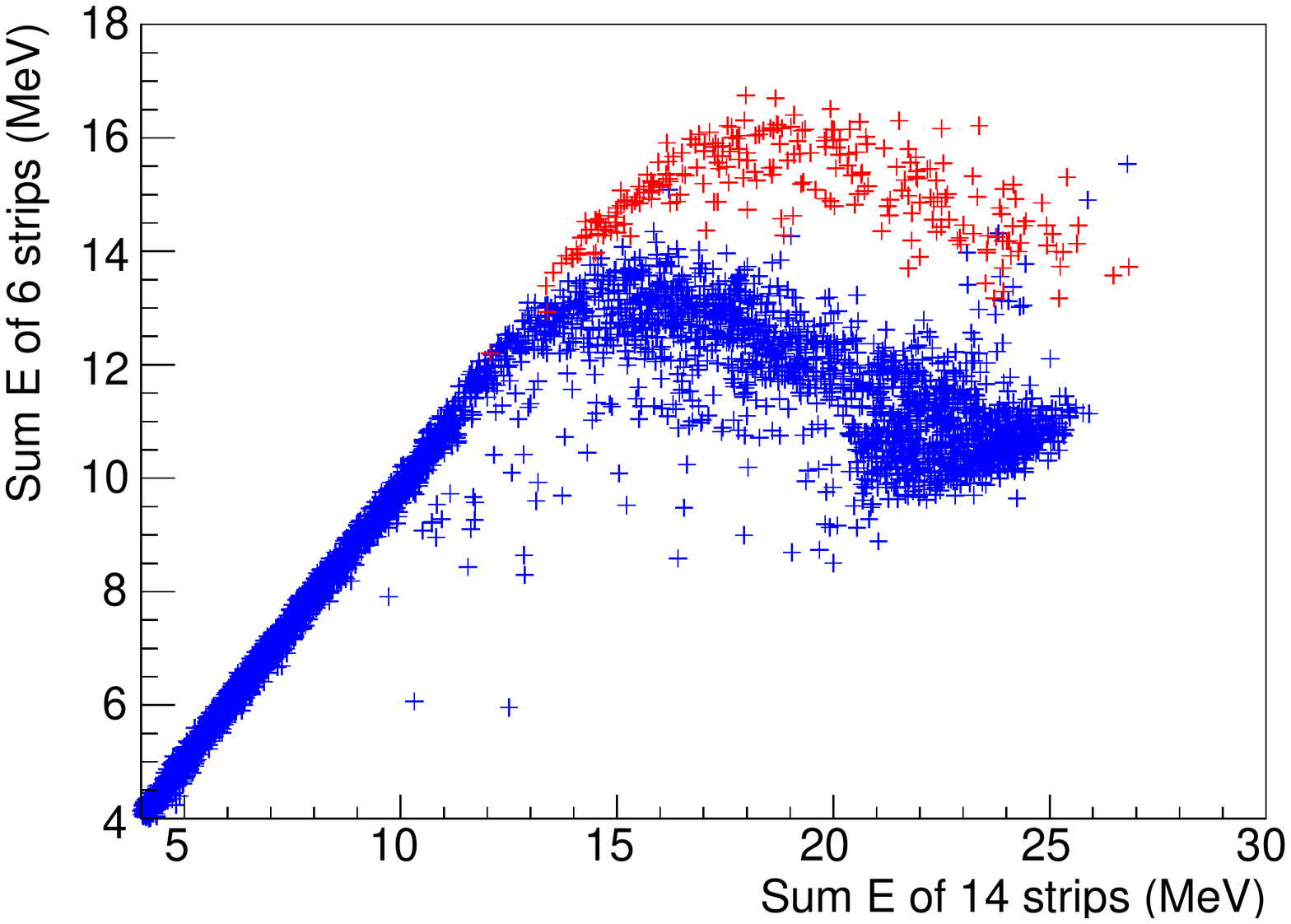}  
  \caption{Strip 2}
  \label{fig:sub-first}
\end{subfigure}
\begin{subfigure}{.32\textwidth}
  \centering
  \includegraphics[width=1.0\linewidth]{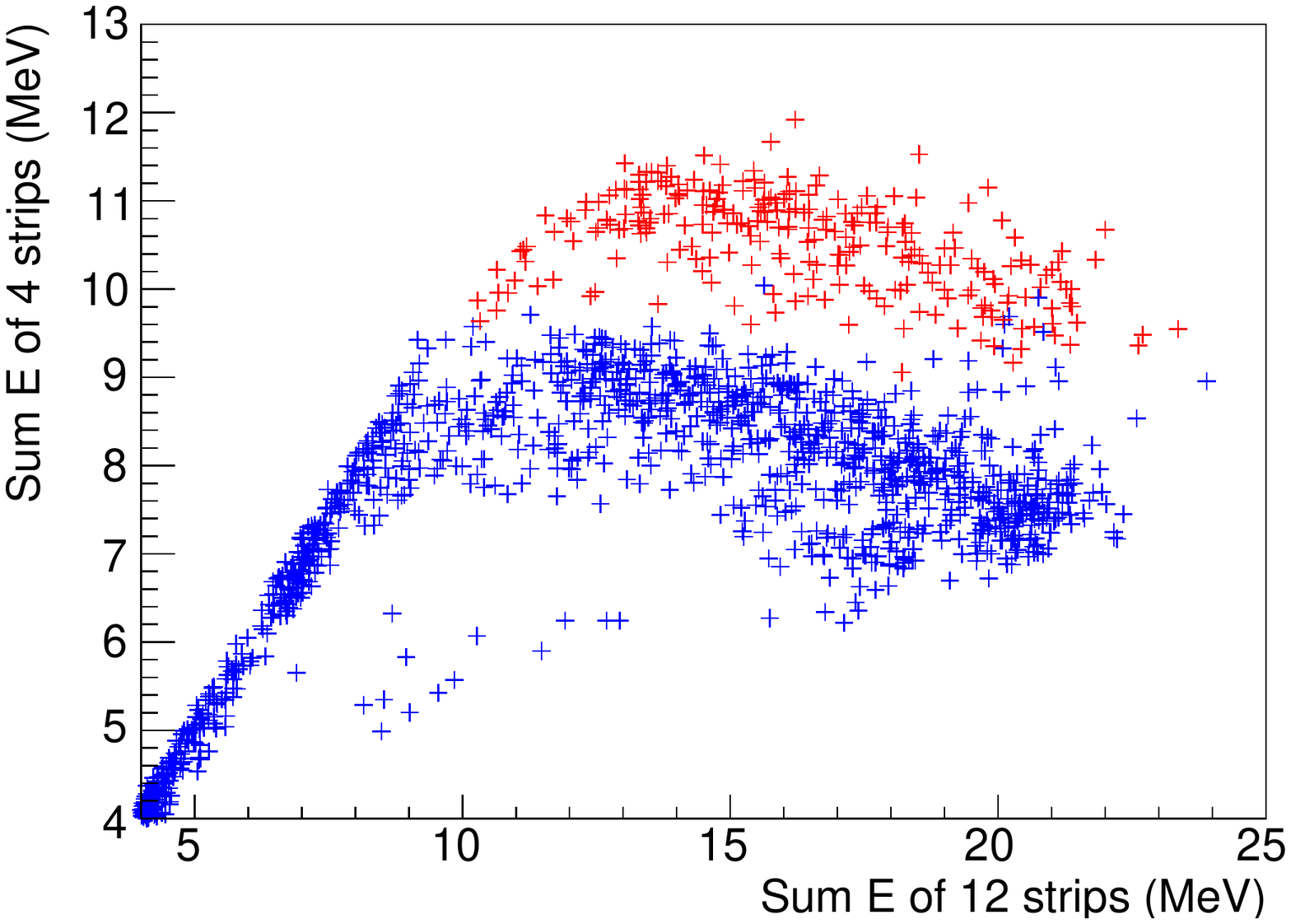}  
  \caption{Strip 4}
  \label{fig:sub-second}
\end{subfigure}
\begin{subfigure}{.32\textwidth}
  \centering
  \includegraphics[width=1.0\linewidth]{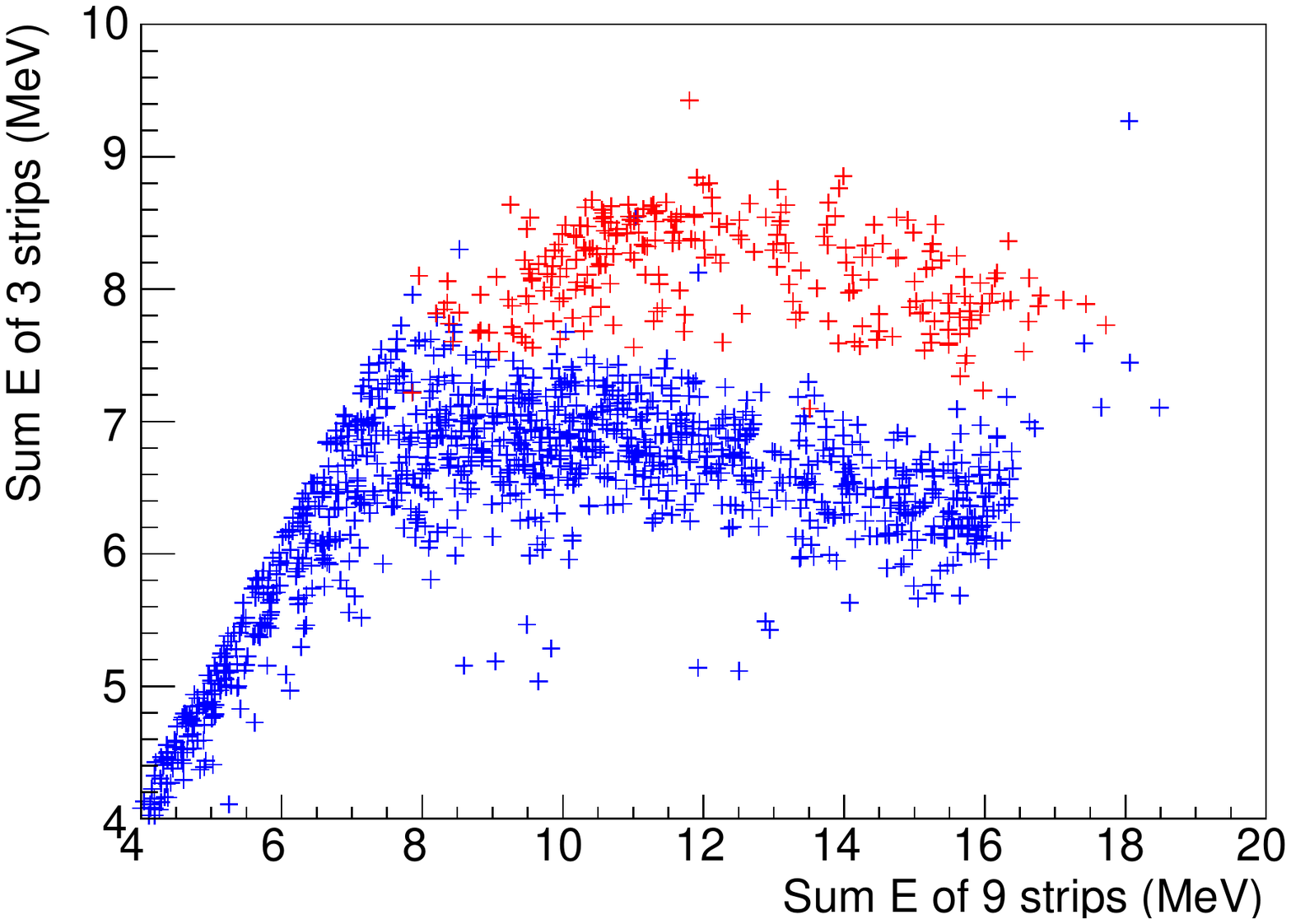}  
  \caption{Strip 7}
  \label{fig:sub-third}
\end{subfigure}
\caption{\label{fig:dEE}$\Delta$E-$\Delta$E plots for MUSIC anode strip 2 (a), strip 4 (b) and strip 7 (c) showing the separation of the \ap events (red points) from \aa events (blue points). The range of strips summed for each $\Delta$E is shown within parentheses.}
\end{figure*}

For the energy range covered in this work, the $(\alpha,p)$, $(\alpha,\gamma)$, and the elastic $(\alpha,\alpha)$ and inelastic $(\alpha,\alpha')$ channels are energetically allowed. Events from the $(\alpha,\gamma)$ reaction are estimated to have cross sections which are 4-5 orders of magnitude lower than the one from the $(\alpha,p)$ reaction. 
To separate out the $(\alpha,p)$ events of interest from the elastic or inelastic events, differences in the amounts of energy deposited in each anode strip are used. As particles move through the detector gas, the energy lost is proportional to the square of the atomic number $Z$ and inversely proportional to the particle energy. When an elastic/inelastic reaction (hereafter denoted as $^{13}$N($\alpha$,$\alpha$)$^{13}$N) or a \Nap reaction occur, due to the Q-values of the reactions and the creation of a heavier nuclei, a "jump" in the energy loss traces can be observed using the signals of the individual anode strips as shown in Fig. \ref{fig:traces}. By summing the energy deposited in various numbers of consecutive strips after a jump in the energy loss occurs allows for the creation of a spectrum for each anode strip where the different reaction channels can be further separated. Fig. \ref{fig:dEE} shows an example of $\Delta E$-$\Delta E$ plots for events occurring in strip 2, strip 4, and strip 7 (red points show \ap events and blue points show \aa events). As part of the data analysis, the selection of events with an incident \N particle was carried out by setting narrow limits on the energy loss signals in the first active strip of MUSIC (strip 0) that encompassed the peak of the energy-loss distribution for \N particles. As seen in Fig. \ref{fig:grid_stp0}, this selection consists of a mixture of the \N beam and contaminants from different charge states of the primary beam.

The total count of the \ap events that occur in each anode strip normalized to the beam intensity provides the total absolute \ap cross section for the relevant center-of-mass energy ($E_{\textrm{c.m.}}$). The present work measures the total \ap reaction cross section within a center-of-mass energy range of 3.26 - 6.02 MeV for anode strips 1 through 9. The energy loss of the \N beam as it travels through the Ti entrance window and the length of the MUSIC detector in the He-Kr gas mixture was calculated using the ATIMA 1.2 energy loss tables \cite{atima} from LISE++ v.13.4.5 \cite{LISE}. This energy loss table was selected because it reproduced the location of the Bragg peak of the energy loss of the \N beam observed using the MUSIC anode pad. Even though the \N beam does not stop at the last anode strip of MUSIC, the Bragg peak can be seen in anode strip 15. The energy loss tables by Ziegler \ea provided in LISE++ \cite{LISE} and SRIM \cite{SRIM} codes show higher stopping powers that is not representative of the experimental data at these beam energies. The uncertainty of the center-of-mass energy covered in each MUSIC strip ranges from 0.19 to 0.24 MeV, and it is dominated by the uncertainty of the laboratory beam energy due to the unknown location of the reaction point (within the width of one strip).

\begin{table}[ht]
\caption{\label{tab:CS} Total reaction cross sections and associated systematic and statistical uncertainties obtained from the present measurement for the \Nap reaction for center-of-mass energies corresponding to anode strips 1-9 of MUSIC. }
\begin{ruledtabular}
\begin{tabular}{ccccc}
\begin{tabular}[c]{@{}c@{}}E$_{\textrm{c.m.}}$\\ (MeV)\end{tabular} &
  \begin{tabular}[c]{@{}c@{}}$\Delta$E$_{\textrm{c.m.}}$\footnote{The energy range per strip is determined by the energy loss of the \N beam along the width of each corresponding strip.}\\ (MeV)\end{tabular} &
  \begin{tabular}[c]{@{}c@{}}$\sigma$\\ (mb)\end{tabular} &
  \begin{tabular}[c]{@{}c@{}}$\Delta\sigma_{sys}$\\ (mb)\end{tabular} &
  \begin{tabular}[c]{@{}c@{}}$\Delta\sigma_{stat}$\\ (mb)\end{tabular} \\
  \hline
6.02 (19) & 0.34 & 347 & 10 & 20 \\
5.70 (20) & 0.28 & 284 & 9  & 18 \\
5.38 (20) & 0.26 & 262 & 8  & 18 \\
5.05 (21) & 0.30 & 306 & 9  & 19 \\
4.72 (21) & 0.32 & 322 & 10  & 19 \\
4.37 (22) & 0.31 & 312 & 9  & 19 \\
4.02 (23) & 0.30 & 304 & 9  & 19 \\
3.65 (23) & 0.21 & 209 & 6  & 33 \\
3.26 (24) & 0.23 & 236 & 67 & 75
\end{tabular}
\end{ruledtabular}
\end{table}

The total reaction cross sections obtained in the present work are shown in Table \ref{tab:CS}, along with the corresponding systematic and statistical uncertainties. The energy binning, $\Delta E_{\textrm{c.m.}}$, represents the estimated energy loss of the beam on a given anode strip. The systematic uncertainty for the calculated cross sections arises from the identification of \ap events from the beam and elastic/inelastic scattering events. The systematic uncertainty was determined by analysing the effect on the number of total \ap events per strip due to different conditions used in the analysis. The uncertainty gradually becomes larger for lower energies (higher anode strip numbers). This is due to the fact that the separation between \ap and \aa events becomes more difficult the closer it gets to the Bragg peak in the energy loss of the beam. The systematic uncertainty for the total cross section is roughly about 3\% for $E_{\textrm{c.m.}}>$ 3.5 MeV, and $\approx$ 29\% for the lowest energy point. The combined statistical and systematic uncertainties (see Table \ref{tab:CS}) are $\approx$ 7\% for $E_{\textrm{c.m.}}>$ 4 MeV, and increases for the lower energies.

\begin{figure}[!ht]
\includegraphics[width=0.5\textwidth]{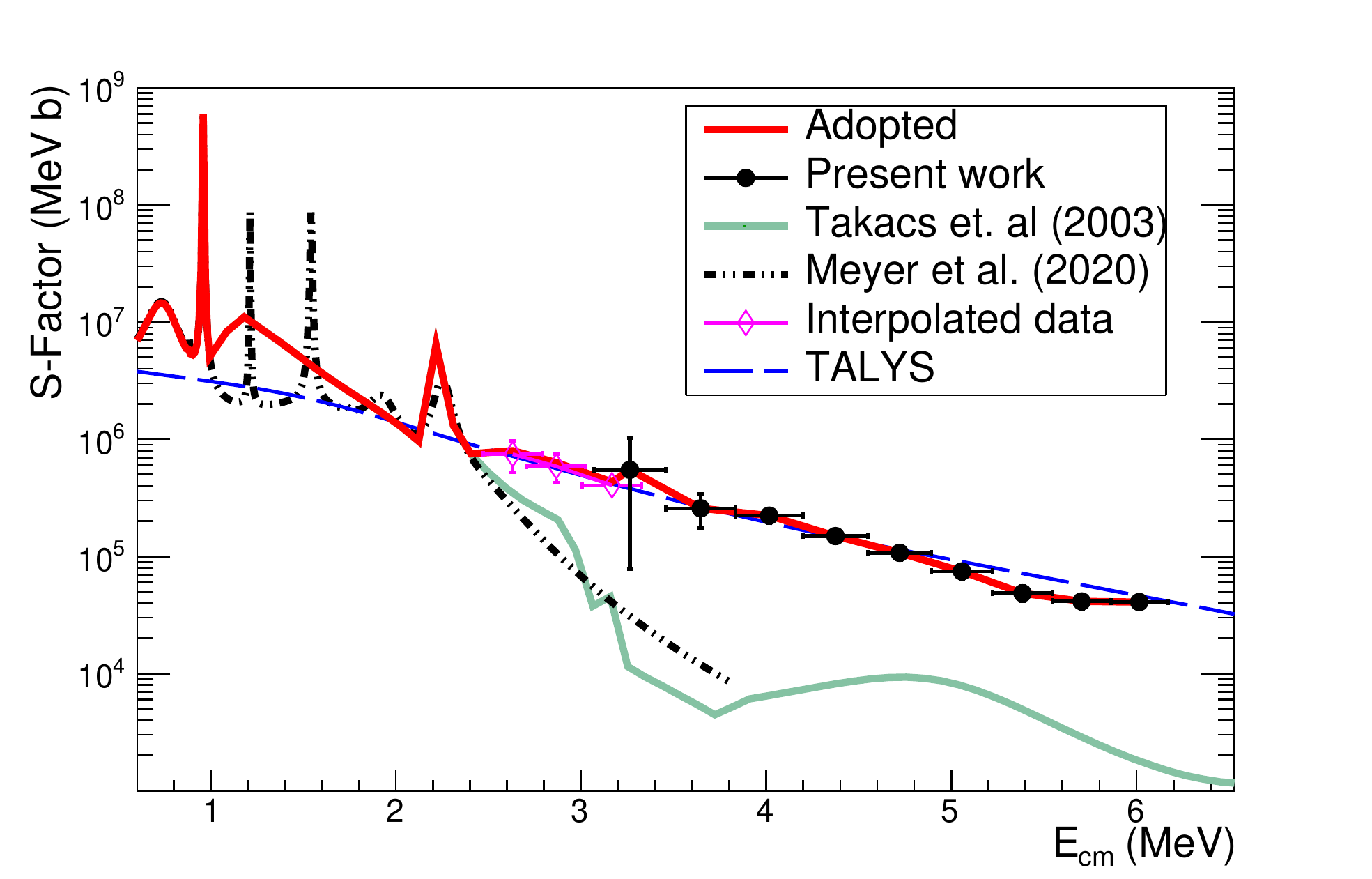}
\caption{\label{fig:SF} Adopted astrophysical S-factors (red) as a combination of the present measurement (black circles), direct \Opa data (green line) \cite{Takacs}, and low energy data from Ref.~\cite{Meyer2020} (black dotted-dashed line). The theoretical cross sections using TALYS are shown by the dashed blue line.}
\end{figure}

There have been several experimental measurements of the \Opa reaction and Takacs \ea \cite{Takacs} provides a fit to several of these \Opa data sets that are currently available. The \Opa cross sections from the fit of Ref.~\cite{Takacs} have been converted to $^{13}$N($\alpha$,p$_0$)$^{16}$O cross sections using detailed balance which gives the reaction cross section for populating the ground state of $^{16}$O. Fig. \ref{fig:SF} shows the S-factor obtained from the present work (black solid circles), the ($\alpha$,p$_0$) compilation (green line), the indirect measurement from Ref.~\cite{Meyer2020} (black dashed-dotted line) and TALYS \cite{TALYS} (blue dashed line). For the TALYS calculations, the McFadden/Satchler alpha optical model potential has been used, which has been shown to be the best at reproducing the reaction cross sections for the mass range $A = 20-50$ \cite{Mohr2015} and below, including the $^{13}$C($\alpha$,$n$)$^{16}$O mirror reaction \cite{Mohr2018,Mohr2017}. 

\begin{figure}[h]
\includegraphics[width=0.5\textwidth]{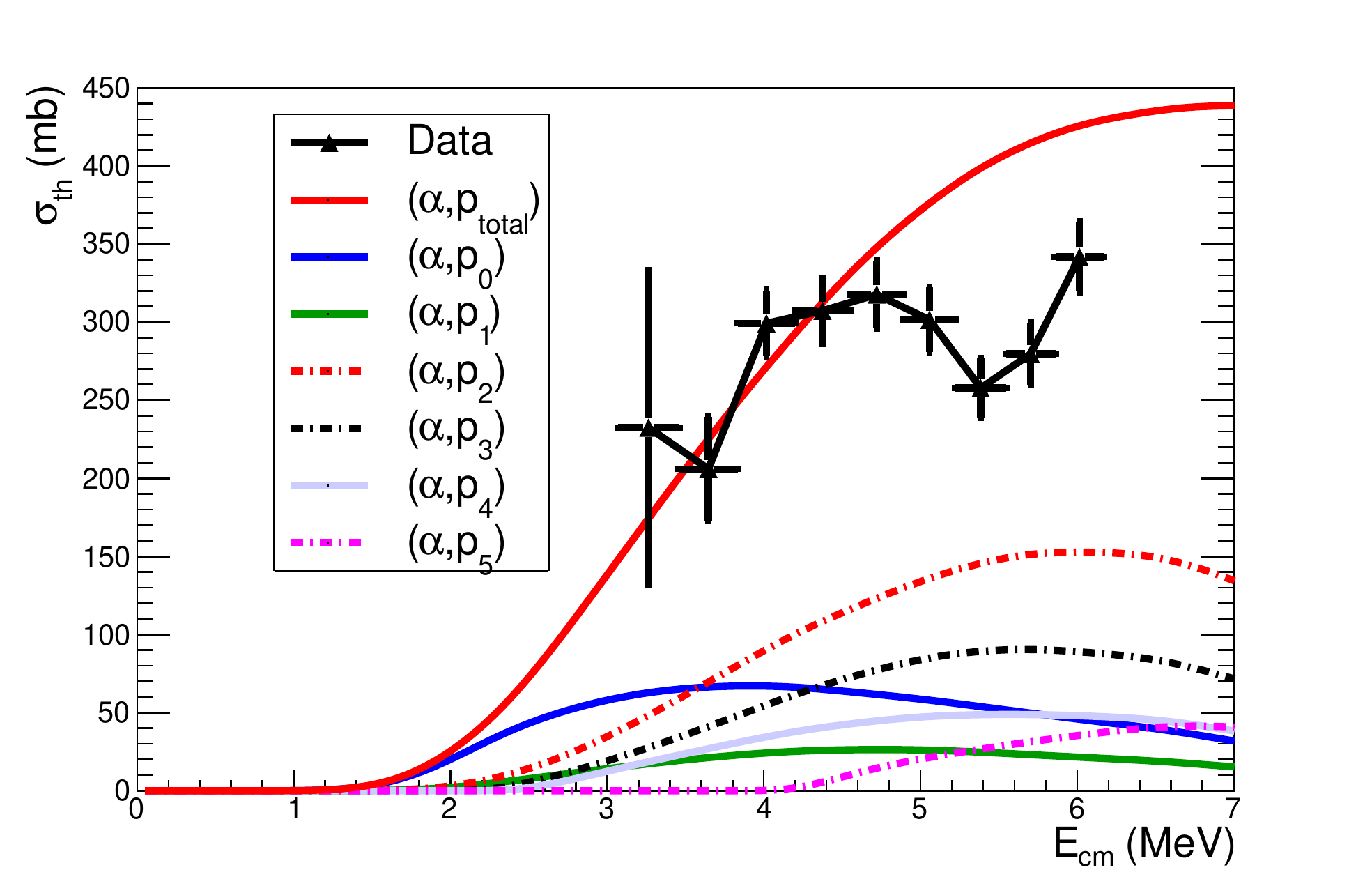}
\caption{\label{fig:exit} Theoretical total cross sections populating different excited levels of $^{16}$O along with the total cross sections from the present work.}
\end{figure}

As can be seen from the astrophysical S-factors in Fig. \ref{fig:SF}, for energies above $\approx$ 2.5 MeV the $^{13}$N($\alpha$,p$_0$)$^{16}$O channel is only a minor contributor ($<$10\%) to the total \Nap reaction cross section. This is mainly due to the dominance of the contributions from higher-lying states of $^{16}$O towards the total \Nap cross section for E$_{\textrm{c.m.}}>$2.5 MeV. The \Nap cross section obtained in the present work is a measurement of the total \ap cross section and includes these contributions from higher exited states of $^{16}$O. The contributions for the total \Nap cross section from the population of different states of $^{16}$O calculated using the statistical model are shown in Fig. \ref{fig:exit}. If the states populated in $^{16}$O are above the $\alpha$-threshold (due to a broad resonance in $^{17}$F which preferentially decays by proton emission to an $\alpha$-unbound state in $^{16}$O), these could in turn decay into $^{12}$C+$\alpha$. Such events where $^{16}$O decays into $^{12}$C+$\alpha$ are not identifiable in the MUSIC detector and, thus, these contributions are missed. Such individual resonances are not included in the statistical model in general, which results in larger theoretical predictions for the \Nap total cross sections when compared to experimental values for E$_{\textrm{c.m.}}\gtrsim$ 5 MeV.

The measured total cross sections from the present work in the energy range of 3.26 - 6.02 MeV allows for the extraction of an accurate reaction rate at temperatures larger than 4 GK. To obtain a comprehensive reaction rate at lower temperatures, previous data from the \Opa reaction, as well as indirect data from Meyer \ea \cite{Meyer2020} have been used. The data from Ref.~\cite{Takacs} results from a fit to several previous \Opa data sets, many of which have been measured at lower energies (1.1 - 2.4 MeV) than in the present work. Thus, the fit from Ref.~\cite{Takacs} was used for energies below 2.4 MeV where mainly the contribution from the ground state is expected, and an uncertainty of 20\% was adopted. For the energies between 2.4 and 3.2 MeV, a few interpolated cross section data points were added, guided by TALYS calculations. This was to ensure a smooth transition between the data from the present work and the $^{13}$N($\alpha$,p$_0$)$^{16}$O data from Ref.~\cite{Takacs} matching the energy resolution obtained in the present measurement. These added data points are shown by magenta open diamonds in Fig. \ref{fig:SF}. For these points an error of 20\% in the cross section was also used. The \Opa data from Ref.~\cite{Takacs} only extends down to E$_{\textrm{c.m.}}\sim$1 MeV in the $^{13}$N+$\alpha$ system. For center-of-mass energies less than 1 MeV, the astrophysical S-factors provided by Ref.~\cite{Meyer2020} have been used. In the energy range of $\approx$ 1-1.8 MeV, two large resonances have been observed in Ref.~\cite{Meyer2020} that were not observed in the compilation of Ref.~\cite{Takacs}, and hence are not included in the present work.

\begin{figure}[!ht]
\includegraphics[width=0.5\textwidth]{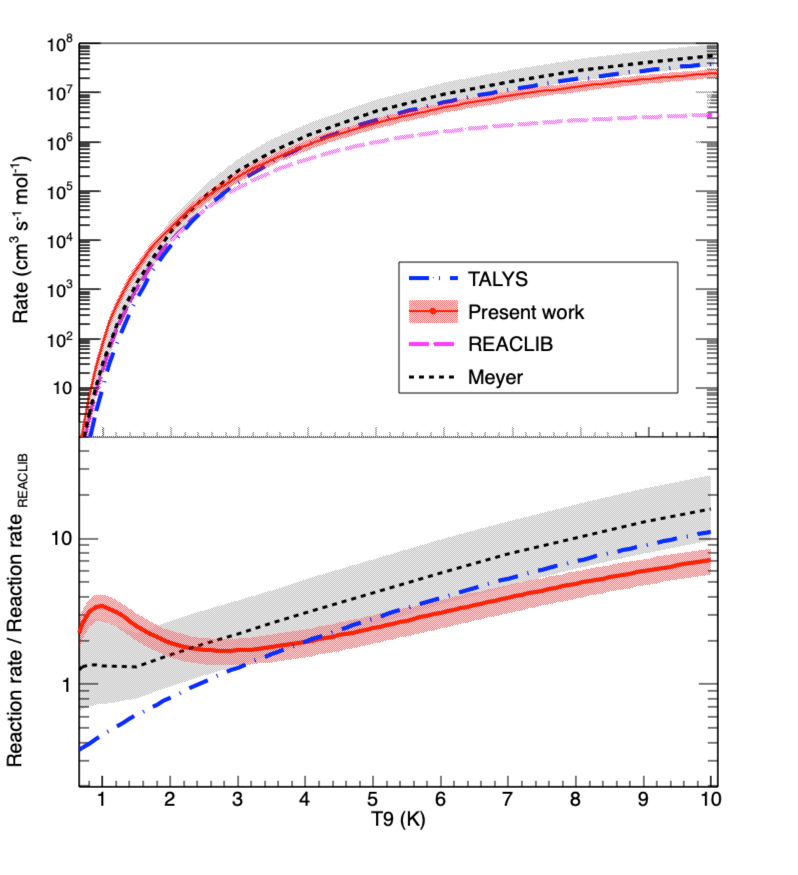}
\caption{\label{fig:rate} The \Nap reaction rate based on this work in comparison to the rates using TALYS, Ref.~\cite{Meyer2020} and REACLIB \cite{Reaclib} (upper panel). The same reaction rates as a ratio to the rate from the REACLIB (lower panel).}
\end{figure}


The resulting astrophysical reaction rate calculated using the code Exp2Rate \cite{Rauscher} is shown in Fig. \ref{fig:rate}. The \Nap reaction rate calculations obtained from TALYS is shown by the blue dashed line along with the reaction rate calculated by Ref.~\cite{Meyer2020} using an indirect method. 

The new adopted rate for the \Nap reaction using the results from the present measurement is higher than the reaction rate from REACLIB obtained with the time-inverse reaction by up to a factor of 3 at temperatures relevant for CCSN. This is not surprising considering that the rate from REALCIB only takes into account the contribution from the ground state. On the contrary, our rate is lower than the rate from Meyer et al. by up to a factor of 3 for temperatures between 2-6 GK. When compared to the rate calculated with TALYS, this is in agreement for temperatures of 3-6 GK (within uncertainties). The discrepancies at higher temperature are probably due to the fact the rate from the present work does not include the contributions from events that decay into $\alpha$+$^{12}$C while TALYS does. The reaction rate calculated using TALYS is significantly lower for T$<$3 GK mainly due to the absence of the strong resonances observed in the compound nucleus $^{17}$F at the lower center-of-mass energies. Overall the reaction rate uncertainties in this work have been reduced dramatically, down to about 20\% at temperatures above 3 GK, which is within the temperature range relevant for the production of the heavy elements in CCSN. More work is needed in the future to more accurately constrain the \Nap reaction rate for the lower temperatures, which is relevant to explain $^{13}$C abundances in presolar SiC grains from CCSN. Model calculations to assess the impact of this new higher reaction rate on the production of heavy elements on CCSN such as $^{44}$Ti and $^{56}$Ni is beyond the scope of this work. However, based on the work by Ref.~\cite{Hermansen2020}, it is estimated that this will increase the production of the heavy elements in CCSN by about a few percent.


To summarize, the first direct measurement of the \Nap reaction has been performed at Argonne National Laboratory using the MUSIC detector. The total reaction cross section has been measured for center-of-mass energies between 3.26 and 6.02 MeV. The new experimental data presented in this work in combination with previous measurements have been used to obtain the \Nap reaction rate. It is found that the adopted rate is lower than that from Meyer \ea by a factor of 2-3, and higher than the REACLIB rate by up to a factor of 3 for temperatures lower than 6 GK, which are relevant for CCSN nucleosynthesis. The rate obtained in this work is in reasonable agreement with the calculated rate using TALYS. More efforts are needed to constrain the reaction rate for the lower temperatures below 2.5 GK.

This material is based upon work supported by the U.S. Department of Energy, Office of Science, Office of Nuclear Physics, under contract number DE-AC02-06CH11357, and National Research Development and Innovation Office (NKFIH), Budapest, Hungary (K134197). This research used resources of Argonne National Laboratory's ATLAS facility, which is a DOE Office of Science User Facility.

\normalem
\bibliography{apssamp}

\end{document}